\documentclass[submission,copyright,creativecommons]{eptcs}
\providecommand{\event}{PXTP 2017} 
\usepackage{breakurl}             
\usepackage{underscore}           
\usepackage{xcolor}
\usepackage{xspace}
\usepackage{proof}
\usepackage{listings}
\usepackage{graphicx}
\usepackage{appendix}
\usepackage{float}
\usepackage{bm}
\usepackage{bussproofs}
\graphicspath{ {images/} }

\newcommand{\blue}[1]{{\color[rgb]{0,0,1} \bm{#1}}}
\newcommand{\fpc}{FPC\xspace}

\newcommand{\pcert}{ProofCert\xspace}

\newcommand{\nnot}[1]{\texttt{not}(#1)}
\newcommand{\nnegate}[1]{\texttt{negate}(#1)}
\newcommand{\ipos}[1]{\texttt{isPositive}(#1)}
\newcommand{\ineg}[1]{\texttt{isNegative}(#1)}
\newcommand{\select}[3]{\texttt{select}(#1,#2,#3)}
\newcommand{\member}[2]{\texttt{member}(#1,#2)}
\newcommand{\ffoc}[1]{\texttt{foc}(#1)}
\newcommand{\uunfk}[1]{\texttt{unfk}(#1)}
\newcommand{\sstore}[2]{\texttt{store}(#1,#2)}
\newcommand{\aand}[2]{\texttt{and}(#1,#2)}
\newcommand{\oor}[2]{\texttt{or}(#1,#2)}
\newcommand{\chain}[3]{\texttt{chain}(#1,#2,#3)}

\newcommand{\chains}[1]{\texttt{chains}(#1)}
\newcommand{\ccheck}[3]{\texttt{check}(#1,#2,#3)}
\newcommand{\certRight}[2]{\texttt{certRight}(#1,#2)}
\newcommand{\certLeft}[2]{\texttt{certLeft}(#1,#2)}
\newcommand{\cute}[4]{\texttt{cuts}(#1,#2,#3,#4)}

\newcommand\proofsystem[1]{\mbox{\slshape #1}\xspace}

\newcommand\LKF  {\proofsystem{LKF}}
\newcommand\PLKF  {\proofsystem{PLKF}}
\newcommand\aLKF {\hbox{\proofsystem{LKF}\kern-2pt$^a$}\xspace}

\newcommand\aLJF {\hbox{\proofsystem{LJF}\kern-2pt$^a$}\xspace}

\newcommand{\Async}[3]{#1\async{#2}{#3}}  
\newcommand{\Sync }[3]{#1\sync{#2}{#3}}  

\newcommand{\wedgep}{\wedge^{\!+}}
\newcommand{\wedgen}{\wedge^{\!-}}
\newcommand{\veep}{\vee^{\!+}}
\newcommand{\veen}{\vee^{\!-}}

\newcommand{\andClerk}[3]{{\wedge_c}(#1,#2,#3)}
\newcommand{\falseClerk}[2]{f_c(#1,#2)}
\newcommand{\orClerk}[2]{{\vee_c}(#1,#2)}

\newcommand{\storeClerk}[3]{\hbox{\sl store}_c(#1,#2,#3)}

\newcommand{\trueExpert }[1]{{\true_e}(#1)}
\newcommand{\andExpert}[3]{{\wedge_e}(#1,#2,#3)}
\newcommand{\andExpertLJF}[6]{{\wedge_e}(#1,#2,#3,#4,#5,#6)}
\newcommand{\orExpert  }[3]{{\vee_e}(#1,#2,#3)}

\newcommand{\initExpert}[2]{\hbox{\sl init}E(#1,#2)}
\newcommand{\cutExpert}[4]{\hbox{\sl cut}_e(#1,#2,#3,#4)}
\newcommand{\decideExpert}[3]{\hbox{\sl decide}_e(#1,#2,#3)}
\newcommand{\releaseExpert}[2]{\hbox{\sl release}_e(#1,#2)}

\newcommand{\tupp}[2]{\blue{\langle #1,}#2{\blue{\rangle}}}

\newcommand{\mkpos}[1]{\partial\kern -1pt_{\scriptscriptstyle +}\kern -1pt(#1)}
\newcommand{\mkneg}[1]{\partial\kern -1pt_{\scriptscriptstyle -}\kern -1pt(#1)}

\title{Determinism in the Certification of UNSAT Proofs}
\author{
Tomer Libal
\institute{Inria Paris\\ Paris, France}
\and
  Xaviera Steele
\institute{American University of Paris\\ Paris, France}
}
\def\titlerunning{Certification of UNSAT Proofs}
\def\authorrunning{T. Libal \& X. Steele}
\begin{document}
\maketitle
\begin{abstract}
   The search for increased trustworthiness of SAT solvers is very active and uses various methods.
  Some of these methods obtain a proof from the provers then check it, normally
  by replicating the search based on the proof's information.
  Because the certification process involves another
  nontrivial
  proof search, the trust we can place in it is decreased.
  Some attempts to amend this use certifiers which have been
  verified by proofs assistants such as Isabelle/HOL and Coq.
  Our approach is different because it is based on an extremely simplified certifier. This certifier enjoys a very high level of trust
  but is very inefficient. In this paper, we experiment with this approach and conclude that by placing some restrictions on
  the formats, one can mostly eliminate the need for search and in principle, can certify proofs of arbitrary size.
\end{abstract}

\section{Introduction}
\label{sec:intro}
SAT solvers feature prominently in many fields of Computer Science, such as Formal Methods \cite{biere1999symbolic},
Security \cite{massacci2000logical}
and Artificial Intelligence \cite{kautz1996pushing}.
It is no surprise that in addition to improving SAT solver performance, the correctness and trustability of their results is
also an area of focus.
One of the main venues that drive and compare
the different solvers
are the annual SAT competitions
 \footnote{\url{http://www.satcompetition.org/}}.

The results of these competitions show a steady
increase in performance of SAT solvers. This improvement is achieved by designing more sophisticated solvers.
However, improvements in design come at a price. The more sophisticated the prover is, the harder it is to understand,
prove its code as correct, and trust its result.
An example of this tradeoff is Lingeling \cite{biere2013lingeling},
a leading SAT solver used extensively in the industry, which was found to have a bug after years of use \cite{wetzler2014drat}.

This lower level of trust prompted research in proof certification which was supported by the community. Since 2013,
producing certificates for checking is mandatory for the main tracks in the SAT competitions. These certificates vary from
extended resolution proofs, like Trace \cite{biere2015}, to formats specifically targeting solvers based on conflict driven clause learning,
such as the DRAT format \cite{wetzler2014drat}.

In order to verify that these certificates indeed represent proofs of a certain theorem, one can build dedicated verifiers
as done by TraceChecker \cite{biere2015} and DRAT \cite{wetzler2014drat}.
These verifiers, while usually simpler than SAT solvers, are based still on non-elementary programs. The possibility of bugs in these
programs reduces the trust we can place in their verification.

One way of amending the problem is to use certified tools to verify proof certificates. Two examples are the verifiers certified
by the Coq \cite{jouannaud2010certification} and Isabelle/HOL \cite{lammich17} proof assistants.

In this paper, we take another approach. Following the ideas presented in the ProofCert project \cite{erc}, we would like
the trusted core of the certifier to be simple to understand and be reproducible by programmers of any skill.
On the other hand, we would like the certification time to be tightly bound to the complexity of the proof (in the way discussed in Sec. \ref{sec:prelim-trace}).

ProofCert certifiers enjoy a high level of trust due to two factors.
The first factor relates to the way they are implemented.
Their trusted
kernel consists of only a few dozen lines of easily-understood Prolog code. Prolog was chosen because the interpreter
takes care of several of the required mechanisms, such as substitutions, unification and backtracking. The simplicity
of the code means that many programmers are capable of writing such certifiers. While writing these certifiers in other
programming languages will require the implementation of the mechanisms mentioned above, the kernels will still be relative simple.
In addition, there already exist many implementations of Prolog which can execute the above certifiers. If a bug exists in one implementation,
we just need to certify the proof evidences using different implementations.

Second, our certifier is using as a reference the propositional sequent calculus, which is one of the most foundational calculi. Compared
to the resolution and other calculi mentioned in this paper, its meta-theory depends on just a few notions such as substitutions and regularity.

ProofCert certifiers also enjoy a high level of universality. Due to the low and foundational level of the sequent calculus, the same
kernel can be used to certify proofs containing different theories, as was done in \cite{chihani2015proof}. This makes an extension of the method
to SMT solvers theoretically possible.

Having a simpler kernel, however, comes with a price: more information than is currently given must be supplied in the proof certificate in order
for the checker to efficiently check a proof.
Taking into account the huge size of unsatisfiability certificates, a further increase in size is clearly not an option.

Fortunately, we find that the information required does not increase the size of the Trace format mentioned above. Since this
format is extractable from other formats, like DRAT-trim \cite{wetzler2014drat}, it can also target generic SAT solvers.

There are different ways for supplementing the trace certificates with the missing information.
The easiest approach would be to preprocess the certificates using a (untrusted) theorem prover.

The key ideas in this paper are using a simple and readable trusted kernel based on the propositional sequent calculus \cite{Gentzen1935}
and defining the certifier so as to guide proof search in an ``almost" deterministic way, which will preserve
a tight bound on the complexity of the certification process.

The prototype certifier presented in this paper targets certificates in the Trace format and can currently certify only
small examples due to an inefficient implementation based on lists, instead of on Prolog's own predicate database.

The implementation is used in order to show that by supplying specific instances of Trace certificates, we can eliminate
most search when looking for a proof over the sequent calculus. Relaxing the restriction on the certificates,
our current certifier's performance declines sharply. We plan on improving its performance by using more efficient data structures.

The approach most similar to ours is the one taken by the first-order resolution certifier ``Checkers" \cite{chihani2015proof}.
The main differences lie in the emphasis on deterministic search in our case as well as in the underlining programming language.

The next section introduces the main technicalities used in the paper, the focused sequent calculus, the ProofCert
approach and the Trace format. We then describe in detail the certifier and its implementation.
The third section will be dedicated to the experiments we have conducted to determine potential relevance for
 very large proofs. We finish with a conclusion and future work section.

\section{Preliminaries}
\label{sec:prelim}
\subsection{Propositional Focused Sequent Calclulus}
\label{sec:lkf}

Theorem provers usually employ efficient proof calculi with a lower degree of trust. At the same time, traditional proof calculi like
sequent calculus enjoy a high degree of trust but are very inefficient for proof search.
In order to use the sequent calculus as the basis of automated deduction, much more structure within proofs needs to be established.
Focused sequent calculi, first introduced by Andreoli \cite{andreoli1992logic} for linear logic,
combine sequent calculi's higher degree of trust with a more efficient proof search. They take advantage of the fact that
some of the rules are ``invertible'', i.e. can be applied without requiring backtracking, and that some other rules can ``focus''
on the same formula for a batch of deduction steps. In this paper, we will make use of the
propositional fragment of the focused sequent calculus system (\LKF) for classical logic defined in \cite{LiaMil09}. Fig. \ref{fig:lkf} presents,
in black font,
 the rules of this fragment (\PLKF).

Formulas in \PLKF\ can have either positive or negative polarity and are constructed from atomic
formulas, whose polarity has to be assigned, and from logical connectives whose polarity is pre-assigned.
The connectives $\wedge^-$ and $\vee^-$ are of negative polarity, while $\wedge^+$ and $\vee^+$ are of positive polarity.

Deductions in \PLKF\ are made during invertible or focused phases. Invertible phases correspond to the application of invertible rules
to negative formulas while a focused phase corresponds to the application of focused rules to a specific, focused, positive formula.
Phases can be changed by the application of structural rules.

In the next section, we will describe a method for certifying proofs of unsatisfiability (UNSAT) using \PLKF.
To make use of the evidence's information while conducting a proof search in \PLKF, we augment
the proof system by adding additional predicates
as seen (in bold, blue font) in Fig. \ref{fig:lkf}. These additional predicates serve only to restrict the proof search in \PLKF\
and, therefore, do not impair the soundness and trustiness of the certifier. The new proof space is a subset of the proof space of \PLKF.
The intuition behind them is that
given a specific UNSAT proof format, we will pair it with a logic program which interprets its meaning
over the focused sequent calculus. For example, in the ''cut'' rule, the implementation of the predicate $\cutExpert{\Xi}{\Xi'}{\Xi''}{B}$
will use the proof evidence, denoted by $\Xi$, to choose the cut formula $B$ which will be then used in \PLKF.
The two remaining components are the proof evidences necessary for the two proofs used by the cut rule.

Additionally, we include a mechanism, in bold blue, to store
formulas by mapping them to indices. This allows for efficient application of the store and decide rules.

\begin{figure}[tb]
\renewcommand{\Async}[3]{\blue{#1}\vdash#2\mathbin{\Uparrow}   #3}
\renewcommand{\Sync }[3]{\blue{#1}\vdash#2\mathbin{\Downarrow} #3}
\renewcommand{\andClerk}[3]{\blue{{\hbox{andNeg$_c$}}(#1,#2,#3)}}
\renewcommand{\falseClerk}[2]{\blue{\hbox{f$_c$}(#1,#2)}}
\renewcommand{\orClerk}[2]{\blue{{\hbox{orNeg$_c$}}(#1,#2)}}
\renewcommand{\storeClerk}[4]{\blue{\hbox{\sl store$_c$}(#1,#2,#3,#4)}}
\renewcommand{\trueExpert }[1]{\blue{{\hbox{true$_e$}}(#1)}}
\renewcommand{\andExpert}[3]{\blue{{\hbox{andPos$_e$}}(#1,#2,#3)}}
\renewcommand{\andExpertLJF}[6]{\blue{{\hbox{andPos$_e$}}(#1,#2,#3,#4,#5,#6)}}
\renewcommand{\orExpert  }[3]{\blue{{\hbox{orPos$_e$}}(#1,#2,#3)}}
\renewcommand{\initExpert}[2]{\blue{\hbox{\sl initial$_e$}(#1,#2)}}
\renewcommand{\cutExpert}[4]{\blue{\hbox{\sl cut$_e$}(#1,#2,#3,#4)}}
\renewcommand{\decideExpert}[3]{\blue{\hbox{\sl decide$_e$}(#1,#2,#3)}}
\renewcommand{\releaseExpert}[2]{\blue{\hbox{\sl release$_e$}(#1,#2)}}

{\sc Invertible Rules}
\[
\infer{\Async{\Xi}{\Theta}{A\wedgen B,\Gamma}}
      {\Async{\Xi'}{\Theta}{A,\Gamma} \quad
       \Async{\Xi''}{\Theta}{B,\Gamma} \quad
       \andClerk{\Xi}{\Xi'}{\Xi''}}
\qquad
\infer{\Async{\Xi}{\Theta}{ A\veen B,\Gamma}}
      {\Async{\Xi'}{\Theta}{ A,B,\Gamma}\quad\orClerk{\Xi}{\Xi'}}
\]
{\sc Focused Rules}
\[
\infer{\Sync{\Xi}{\Theta}{B_1\wedgep B_2}}
      {\Sync{\Xi'}{\Theta}{B_1}\quad
       \Sync{\Xi''}{\Theta}{B_2}\quad
       \andExpert{\Xi}{\Xi'}{\Xi''}}
\qquad
\infer{\Sync{\Xi}{\Theta}{B_1\veep B_2}}{\Sync{\Xi'}{\Theta}{B_i}\qquad
       \orExpert{\Xi}{\Xi'}{i}}
\]
{\sc Identity rules}
\[
\infer[cut]{\Async{\Xi}{\Theta}{\cdot}}
           {\Async{\Xi'}{\Theta}{B}\quad
            \Async{\Xi''}{\Theta}{\neg{B}}
            \quad \cutExpert{\Xi}{\Xi'}{\Xi''}{B}}
\qquad
\infer[init]{\Sync{\Xi}{\Theta}{P_a}}
            {\tupp{l}{\neg P_a}\in\Theta\quad\initExpert{\Xi}{l}}
\]
{\sc Structural rules}
\[
\infer[\kern -1pt release]{\Sync{\Xi}{\Theta}{N}}
               {\Async{\Xi'}{\Theta}{N}\quad\releaseExpert{\Xi}{\Xi'}}
\qquad
\infer[store]{\Async{\Xi}{\Theta}{C,\Gamma}}
             {\Async{\Xi'}{\Theta, \tupp{l}{C}}{\Gamma} \quad
              \storeClerk{\Xi}{C}{l}{\Xi'}}
\]
\[
\infer[\kern -1pt decide]{\Async{\Xi}{\Theta}{\cdot}}
              {\Sync{\Xi'}{\Theta}{P}\quad
               \tupp{l}{P}\in\Theta\quad
               \decideExpert{\Xi}{l}{\Xi'}}
\]
\caption{The \PLKF proof system.
  The symbol $P_a$ denotes a positive atomic formula, $N$ a negative formula and $C$ a positive formula or a negative literal.}
\label{fig:lkf}
\end{figure}

\subsection{A General UNSAT Proof Checker}
\label{sec:pcert}

There is no consensus about the optimal shape for a formal proof evidence.
The notion of structural proofs, based
on derivations in some calculus, is of no help as long as the calculus is not fixed.
One of the goals of the \pcert\ project
\cite{erc} is to amend this problem by defining the notion of a foundational proof certificate (\fpc): a pair of some arbitrary
proof evidence and an executable specification that denotes its semantics in terms of some well-known target calculus,
such as the sequent calculus. These two elements of an \fpc\ are then
given to a universal proof checker which, by the help of the \fpc, can derive a proof in the target calculus.
Since the proof generated is over a well-known and low-level calculus that is easy to implement, one can obtain
a high degree of trust in its correctness.

Such a universal proof certifier, which will be mainly implemented
in the programming language Prolog,
contains the following main components.

\begin{itemize}
  \item {\bf Kernel.} The kernels are the implementations in Prolog of several trusted proof calculi. Currently, there
    is a kernel over the propositional classical focused sequent calculus (\PLKF). Section \ref{sec:lkf}
  presents the calculus \PLKF{} that will be used in the paper.
  \item {\bf Proof evidence.} The first component of an \fpc, a proof evidence is a description
    of the proof output by a theorem prover. Given the high-level declarative form of Prolog, the structure and
    form of the evidence are very similar to the original proof.
    We will see the precise form of the different proof evidences we handle in the next section.
  \item {\bf \fpc specification.}
    The basic goal of the universal proof certifier is to generate a proof of the evidence's theorem
    in the target kernel.
     To this end, the kernels have additional predicates
    which take into account the information given in the evidence. Since the form of this information is not known
    to the kernel, the certifier uses \fpc specifications to interpret it. These
    logical specifications are written in Prolog and interface with the kernel in a sound way in order
    to certify proofs. Writing these specifications is the main task for supporting the different outputs of
    the modern theorem provers that we consider in this paper and they are explained in detail in Section \ref{sec:cert}.
\end{itemize}

\subsection{The Trace Proof Format}
\label{sec:prelim-trace}

The description given in the previous sections refers to a universal proof certifier for SAT solvers.
In this paper, though, we focus on certifying a specific proof format and describe the implementation and some experiments.

The annual SAT competition has included a certified UNSAT category since 2005 and has required certifications for the general UNSAT track
since 2013. One of the proof formats which was supported until 2013 is Trace\footnote{\url{http://fmv.jku.at/tracecheck/}}.
Trace is a format supporting resolution proofs where the resolution steps might require arbitrary input resolution proofs.
This flexibility allows Trace to support, on one hand, resolution and hyper resolution proofs, and derivations based on clause learning on the other.
In order to simplify integration with the SAT solvers that use clause learning, Trace
allows arbitrary orders between components in the proof. Since 2014, the SAT solver community has preferred to use formats
which are specialized for learned clauses. These formats are easier to produce for these solvers as well as are more
compact in size. However, these formats are harder to certify and require elements of theorem proving.

In this paper, we follow the \pcert\ approach of requiring the trusted kernel of a proof certifier to be simple so it can be implemented
by programmers who have no knowledge of theorem proving and only basic knowledge in logic. To this end, we chose to support Trace as
its format best serves this goal.
Although it is not supported by SAT competitions, a proof in Trace
format can be extracted from a DRAT \cite{wetzler2014drat} proof, the current state-of-the-art UNSAT proof format.

The basic components of the Trace format are the {\em chains}. Chains represent the production of a derived clause from
a set of previously known clauses based on {\em linear input resolution}. This means that the resolution proof is, in fact,
of the shape of a tree and can be linearly (with respect to the number of cuts) converted into a sequent calculus proof.
In order to support clause learning, the known clauses as well as the chains themselves
can be supplied in any order.

A chain contains three parts: the index of the derived clause, the literals of the derived clause and a list of antecedents.
The antecedents are both preceded and followed by zero (`0'). For example, to denote the clause with the first variable
occurring positively and the second negatively, with an index `3', and which is derived using clauses from chains `1' and `2',
we would write the following.

\begin{lstlisting}
  3 1 -2 0 1 2 0
\end{lstlisting}

The original clauses are represented using chains with no antecedents as follows

\begin{lstlisting}
  1 1 3 0 0
\end{lstlisting}

The empty clause is denoted by an empty list of literals.

Since the literals of a derived clause can be derived from the list of antecedents, Trace also supports chains without
an explicit list. This is achieved by replacing the list with the `*' symbol.

A simple example of a Trace proof (taken from Trace website \footnote{http://fmv.jku.at/tracecheck/README.tracecheck})
can be seen in Fig. \ref{fig:trace-ex}.

\begin{figure}[tb]
  \begin{lstlisting}
  1 1 2 0 0
  2 -1 2 0 0
  3 1 -2 0 0
  4 -1 -2 0 0

  5 1 0 3 1 0
  6 0 4 2 5 0
  \end{lstlisting}
  \caption{A example of a Trace refutation}
  \label{fig:trace-ex}
\end{figure}

\section{Certification of the Trace Format}
\label{sec:cert}
As mentioned in the introduction, current approaches for certifying SAT solver-generated proofs involve
using the proof evidence to guide proof search using, for example, a resolution calculus.
This approach, while enjoying a higher level of trust than that of the SAT solver alone, suffers from some of the same problems.
Namely, some proof search is required in order for
the certification process to succeed. This essentially means that the certifier is in itself a resolution theorem prover.

The approach taken in this paper is first to use a lower level and foundational calculus, namely the sequent calculus. Second,
we attempt to reduce proof search to nearly deterministic ``search'' only, thus making the certification process more efficient and
separating it from the proof search done by the SAT solver.
Sec. \ref{sec:exps} will be devoted to our experiment with this approach and its applicability to the Trace SAT proof format.
In the rest of this section, we will describe the details of the certifier and how it implements the ideas described in Sec.
\ref{sec:pcert}.

\subsection{Prolog and Proof Search}
For our certifier, the first step was to implement a highly trusted kernel following the \PLKF rules. For this purpose,
we chose the Prolog language. Prolog's unification-based computational model lends
itself naturally to proof search: given some appropriately formatted proof evidence, Prolog attempts to pattern-match
and see if there are any applicable rules. If there is, the rule is applied, and the branch is continued until either
success (application of the ``init'' rule, in this case) or failure (there are no more rules to apply). If the branch fails,
Prolog will automatically backtrack to the last fork (when there was either more than one applicable rule,
or more than one variable the rules could apply to) and test another option. This process creates a proof search very
similar to what a human would manually write using a sequent calculus, except that one would expect the human to more intuitively
identify whether a branch will advance the proof or end in failure.

The focused calculus rules presented in Fig. \ref{fig:lkf} are closely followed by the Prolog code, so as
to maintain soundness and completeness. There are a few adjustments for efficiency, which we will point out in this section;
we will also explain the behavior
and meaning of the Prolog predicates and variables so the reader can easily confirm their faithfulness to the original calculus.
Then we introduce the translation of a Trace proof into Prolog. Finally, we discuss in more detail the implementation
of the ``Cut'' rule, as this is where outside input (the Trace, in this case) is introduced.

\subsection{Syntax}
As we have mentioned, once the rules are written in Prolog, Prolog's internal backtracking takes care of the proof search.
To grasp the Prolog rules and their faithfulness to \PLKF, the reader need only be familiar with a few syntactical
and implementation details. Firstly, words beginning in lower case are either predicates or constants in Prolog, while
those beginning with an upper-case letter are variables that Prolog can try to unify with other variables or constants.
An underscore ( _ ) in Prolog represents an anonymous variable, which are only used in cases where the previous
value of the expected parameter no longer needs to be kept.

Secondly, keep in mind Prolog's list syntax: both $X$ and $[X|Y]$ are variable formats that
can match with a list. The first will match X to the whole list, whereas the second notation assigns $X$ to the first element
of the list, and $Y$ to the remainder. The variable names, of course, may vary as long as the first letter is capitalized.
Lastly, we use a built-in predicate, $\select{X}{A}{A_1}$.
It succeeds if $X$ is a member of list $A$, and $A_1$ is the result of removing $X$ from $A$.
Another built-in predicate,
 $\member{X}{List}$, is self-evident.

\subsection{Kernel Implementation Details}
\label{sec:imp-details}
In our implementation, we use $x(P)$ to
denote an atomic formula and \nnot{$x(P)$} to denote a negative atomic formula. When we need to negate a non-atomic formula
(which only occurs in one rule),
we can use the \nnegate{} predicate. The helper predicates, \ipos{Formula} and \ineg{Formula},
succeed if $Formula$ is positive or negative, respectively.
As mentioned in Sec. \ref{sec:prelim-trace},
$\wedge$ (\texttt{and}) is assigned a positive polarity; $\vee$ (\texttt{or}) a negative polarity.
Atomic formulas are positive by default, and negative when
they are wrapped in the \nnot{} predicate.

When creating a proof in the \PLKF system, one maintains at most two sets of formulas.
One set, denoted by $\Theta$ in Fig. \ref{fig:lkf}, is the context;
its formulas are neither focused nor unfocused.
The other set, depending on the phase, consists of zero or more unfocused formulas, or exactly one focused formula.
In our Prolog code, these sets are implemented as lists. The predicates $\uunfk{}$ and $\ffoc{}$ are used to differentiate
unfocused and focused phases. It is the ``store" rule that adds a formula to the set $\Gamma$.
What is, in \PLKF, the $\Gamma$ set is now divided into two   lists, themselves within the
$\sstore{SL}{NL}$ predicate.

One list, referred to by the $NL$ variable,  contains only negative atomic formulas; $SL$ contains the rest of the stored formulas.
This division, implemented in the ``store" rule, improves the
efficiency of the
  ``init" rule, which would otherwise need to check the whole list for a negative atomic formula.
It has no impact on the soundness or completeness of the system since according to the rules of \PLKF, when
deciding on a stored formula,
negative atoms cannot be chosen. This division then necessitates two versions of the ``store" rule:
one that applies to positive formulas and stores
them in the $SL$, and another that applies to negative atoms and stores them in $NL$. Having two versions does not affect
the kernel since they are mutually exclusive and never conflict or overlap.

The resulting trusted kernel is as follows:

\begin{lstlisting}[language=Prolog]
%true focused, unfocused
check(_,_,foc(true)).
check(_,_,unfk([true|_])).

%false unfocused
check(Cert,Store,unfk([false|Gamma])):-check(Cert,Store,unfk(Gamma)).

%init
check(Cert,store(_,NL),foc(x(P))):-inite(Cert), member(not(x(P)),NL).


%release; N is a negative literal or formula
check(Cert,SL,foc(Formula)) :-
    isNegative(Formula),
    releasee(Cert,Cert1),
    check(Cert1,SL,unfk([Formula])).

%cut
check(Cert,Store,unfk([])) :-
    cute(Cert,Cert1,Cert2,Formula),
    negate(Formula,NFormula),
    check(Cert1,Store,unfk([Formula])),
    check(Cert2,Store,unfk([NFormula])).


%decide
check(Cert,store(SL,NL),unfk([])) :-
    decidee(Cert,Cert1,Index),
    member((Index,Formula),SL), isPositive(Formula),
    check(Cert1,store(SL,NL),foc(Formula)).


%and focused
check(Cert,SL,foc(and(A,B))) :-
    ande(Cert,Cert1,Cert2),
    check(Cert1,SL,foc(A)), check(Cert2,SL,foc(B)).

%or unfocused
check(Cert,SL,unfk([or(A,B)|Gamma])) :-
    ore(Cert,Cert1),
    check(Cert1,SL,unfk([A,B|Gamma])).


%store; negative atom case
check(Cert,store(SL,NL),unfk([not(x(P))|Gamma])) :-
    storee(Cert,Cert1,_),
    check(Cert1,store(SL,[not(x(P))|NL]),unfk(Gamma)).

%store; positive formula case
check(Cert,store(SL,NL),unfk([C|Gamma])) :-
    isPositive(C),
    storee(Cert,Cert1,Index),
    check(Cert1,store([(Index,C)|SL],NL),unfk(Gamma)).
\end{lstlisting}

Except for differences discussed above, these predicates follow precisely the
modified \PLKF from Fig. \ref{fig:lkf}. These rules are evidently concise, making them easier to write and test, so
the likelihood of bugs is diminished. Furthermore, because of the code's brevity, the kernel could also be replicated
in another language, so multiple kernels could be tested on the same problem to check the result. Therefore our
kernel's simplicity is an advantage in terms of current trustability, and could easily support further
improvements in the future.

\subsection{Proof Evidence}

{\bf Resolution Refutation} -
A Trace file succinctly represents an extended resolution refutation proof, in conjunctive normal form (CNF).
In this format, each original clause (as described in \ref{sec:prelim-trace}) in the Trace corresponds to one
clause,
and every line is connected by a conjunction ($\wedge$). A resolution refutation uses two or more such clauses to derive a new clause,
until $false$ can be derived. The original formula used to derive $false$, then, must be unsatisfiable.
If this is the case, the negation of the original formula should be $true$, and this is what
our theorem prover can confirm. To do this, we can simply negate the original formula and pass it to the Prolog theorem prover.

With this process, we would be recreating the work of the SAT solver, just with a different underlying proof search mechanism.
The purpose of our checker, however, lies in its incorporation of the Trace's derived clauses, each of which can
be translated to an instance of the ``cut'' rule.
In this section, we discuss how we translate a Trace's proof evidence into code interpretable by Prolog.

{\bf Proof evidence: format for original clauses} -
A simple Python program integrates the Trace proof into the Prolog kernel syntax. It is a text-to-text
translator, taking the Trace format input and returning output readable by Prolog. The only nontrivial
part is translating from Trace's implicit infix notation to the prefix notation
accepted by our Prolog kernel, which is done recursively.
In this section we explain the
syntax used in Prolog that corresponds to the different elements in a Trace.

The Trace's list of original clauses corresponds to a
single formula in disjunctive normal form (because we have negated the initial formula in order to use resolution
refutation), using the predicates $\aand{F}{F1}$ and $\oor{F}{F1}$. For example, take a single clause from a hypothetical Trace:

\begin{lstlisting}
  7 -1 2 -4 0 0.
\end{lstlisting}

The first integer is the index of the clause, which we will deal with below.
The rest of the clause becomes: $\aand{x(1)}{\aand{\nnot{x(2)}}{x(4)}}$. The trace's Trace's other clauses would be similarly
translated, then they would all be linked by disjunctions using the same method, eventually resulting in a string of the general form: \\
$\oor{\aand{x(1)}{\aand{\nnot{x(2)}}{x(4)}}}{\aand{...}{...}}$. \\
Thus the first part of the proof evidence,
the theorem that should be validated, is passed to Prolog as a string recursively generated from the Trace.
Prolog then
reads the string as a query, which it can try to patternmatch with the previously defined predicates.
Initially, the theorem is unfocused and so the whole formula is wrapped in the \uunfk{} predicate.

To preserve the indices given to clauses in a Trace, there are some other adjustments.
When the ``store" rule is applied, it stores the clauses as a tuple of $(Index, Clause)$ in the $SL$ list,
wrapped in the $\sstore{SL}{NL}$ predicate.
To do this, we add a
variable, $DexList$,
to the initial program call. $DexList$ contains the indices of all the
original (as opposed to derived) clauses and ensures
that when they are stored, the clauses keep the index assigned by
the SAT solver. Because the program begins in the
unfocused phase, and the original formula is a string of clauses in disjunctive normal form, individual formulas will
be progressively broken down by the unfocused ``or" ($\vee$ rule, then stored with the appropriate index by $\sstore{SL}{NL}$.

After this initial unfocused phase, original clauses will all have been stored with the indices assigned them in the Trace.
Subsequent calls to ``store", which will be to store positive literals on the left proof branch, will use the index
$-1$. Because there can be more than one stored positive literal, $-1$ is no longer a unique index. So, deciding on $-1$
can yield any of the literals stored there.
In the next section, we discuss more specifically how the Trace proof format is
mapped to a consistent syntax in Prolog.

{\bf Proof evidence: format for derived clauses} -
The last element of the proof evidence from Trace are the derived clauses or chains, and the indices that
the original clauses need to be stored with. As we have seen in Sec. \ref{sec:prelim-trace},
a chain contains the index, a
derived clause, and a list of the indices of its antecedents. This information is stored in the
predicate $\chain{Index}{DecideList}{Formula}$; then multiple $\chain{I}{D}{F}$ objects are stored in a list, which is wrapped
in a $\chains{...}$ predicate. Here, $Index$ is an integer, $DecideList$ is a Prolog list of integers (the indices of antecedents),
and $Formula$ is a single formula. Whereas the Python Trace-Prolog translation program automatically negates the original
theorem in preparation for resolution refutation, the derived clauses are not negated when translated to Prolog.

If we add some antecedents to our previous Trace clause example, the clause becomes:

\begin{lstlisting}
  7 -1 2 -4 0 5 6 8 0.
\end{lstlisting}

This would be classified as a derived clause thanks to its nonempty antecedent list, and would be written in Prolog as:\\
$\chain{7}{[5, 6, 8]}{\oor{\nnot{x(1)}}{\oor{x(2)}{\nnot{x(4)}}}}$. \\

An entire list of derived clauses would have the following form:

$\chains{[\chain{7}{[5, 6, 8]}{\oor{\nnot{x(1)}}{\oor{x(2)}{\nnot{x(4)}}}},\chain{I}{D}{F},\ldots]}$. \\

{\bf Proof evidence: combining elements to convey an entire Trace in Prolog} -
To summarize, a derived clause in the Trace is negated then directly translated into Prolog.
The original clauses are translated without being negated, while  their indices are stored separately as described above.

The format for calling the verification on a whole Trace is as follows:

$\ccheck{Cert}{Store}{\uunfk{Formula}}$. \\

$Store$ is a predicate of the form $\sstore{SL}{NL}$ as mentioned in Sec. \ref{sec:imp-details},
where $SL$ and $NL$ are both lists of formulas separated by formula type (positive formula or negative atom).
Initially both will be empty.

The $Cert$  variable contains the proof evidence. It is where we keep all information that comes exclusively from the Trace,
that is not found in the original CNF SAT problem formulation. Thus the list of derived clauses we saw above,
\chains{}, is managed in the $Cert$ variable. In the next section, we will give details about how and when $Cert$ is
updated. For now, we simply note that the initial checker call requires that Cert be of the form
$\certRight{IndexList}{\chains{ChainList}}$.
As noted in Sec. \ref{sec:imp-details}, $IndexList$ contains the assigned indices of all original clauses.
In summary, we rearrange the format and the storage of indices with respect to the original Trace proof,
but otherwise the Prolog formulation corresponds directly to the original proof.

\subsection{Incorporation of the Trace: FPC File}

As mentioned above, information that is obtained exclusively from the Trace is only manipulated in the Prolog
checker code through the variable $Cert$. Any operations updating or using information within the $Cert$ variable
are managed not by the \PLKF\ kernel, but by additional predicates within the \fpc file. The
reader may note that while the predicates in \fpc can
guide or restrict the proof search, they cannot create new branches or add elements to the search.
They therefore have no
impact on the soundness of the PLKF system. Setting up the \fpc code in this way
allows us to include the Trace's guidance without impacting soundness.

After an application of the cut rule, the proof search must succeed on both resulting branches. The left cut branch
corresponds to a sub-proof, one that states that the cut formula follows from the clauses whose indices are given in the cut's $DecideList$
variable. If this succeeds, then the checker proceeds to the right branch, which repeats the same process on the remaining chains,
and should be able to confirm that all the chains together form a valid proof. Thus on the left branch, the information needed from
the $Cert$ context is only the $DecideList$. On the right branch, the checker needs the information on all the remaining chains, plus
the index at which to store the current chain's derived clause.

Thus, the left and right branches each require different information from the $Cert$ context. So we divide this variable into
$\certLeft{DecideList}{Flag}$ and $\certRight{IndexList}{\chains{ChainList}}$. This improves the code's readability as well as
efficiency: rules that are only applicable on one of the branches can fail sooner this way. With this syntax of the
$Cert$ variable in mind, we can look at the predicates in the \fpc file.

\begin{lstlisting}[language=Prolog]
% cut expert
cute(certRight([],chains([chain(StoreDex, DL, Formula)|RestChains])),
  certLeft(DL,1),
  certRight([StoreDex],chains(RestChains)),
  Formula).
\end{lstlisting}

The cut expert is called with the following parameters: $\cute{Cert}{Cert1}{Cert2}{Formula}$.
This method makes sure each branch receives, through the $Cert$ variable, the information it needs to proceed.
Then it returns two modified $Cert$ contexts, one corresponding to the left cut branch; one to the right.
We see that this rule is only applicable when the list of unfocused formulas is empty, and when the $Cert$ variable is currently of type
$certRight$.

The left branch's certificate contains only the $DecideList$ and a flag variable, which is initialized to $1$ at the beginning of each left branch.
The flag variable is only utilized in the ``decide" rule, as detailed below.

If the left branch succeeds, the kernel code will continue on the right branch,
this time with information on the remaining chains it must check, and the negated version of the cut formula. Due to the
format of Trace proofs, the negated formula will always be of positive polarity (a conjunctive clause), and thus will be immediately
stored at that index. There are also some cases where the cut formula may be an atomic formula, in which case its negation may be
either positive or negative; $Store$ applies in either case.

\begin{lstlisting}[language=Prolog]
% decide expert
decidee(certLeft(DL,1),certLeft(DL,0),-1).
decidee(certLeft(DL,1),certLeft(DL1,1),I) :- select(I,DL,DL1).
\end{lstlisting}

The decide expert extracts and updates the information in $Cert$, returning an index for the ``decide" rule and
then removing that index from $DecideList$ so it can only be selected once.
In the case of subformulas stored at the non-unique index $-1$, the flag variable is in charge of restricting
the number of decides so that  we can only decide on one of potentially several subformulas per branch.
This is not an issue, as if it is the wrong formula, Prolog can backtrack, swapping  the flag variable to $1$ again,
and choosing a different formula stored with $-1$.
As mentioned earlier, this non-determinism is very shallow.

\begin{lstlisting}[language=Prolog]
% store expert
storee(certRight([Index|Rest],Chains),certRight(Rest,Chains),Index).
storee(certLeft(DL,A),certLeft(DL,A),-1).
\end{lstlisting}

Both these two cases of the store expert may be called by both cases of the kernel's ``store" rule.
If we are on the right branch, we will have index(es) from either the initial unfocused phase,
or from a previous cut introduction; the next formula we store should be with that index, which we remove from the
 $IndexList$.
On the left branch there are no indices from the Trace, and
we are only ever storing literals,
thus the expert makes no change to the DecideList, and returns -1 as the Index to store at.

\begin{lstlisting}[language=Prolog]
% init expert
inite(_).
% release expert
releasee(certLeft(DL,A),certLeft(DL,A)).
% and expert
ande(certLeft(DL,A),certLeft(DL,A),certLeft(DL,A)).
% or expert
ore(Cert,Cert).
\end{lstlisting}

These remaining expert predicates do nothing, and are only included to maintain a uniform syntax.
The kernel rules that call them require neither information nor changes to the Cert.
 ``release" and ``and" experts, though,
do have the advantage of making the proof search slightly more efficient by failing earlier on a wrong branch.
As ``or" can be applicable on both right and left branches, this is left ambiguous.
Finally, if the kernel ``init" rule succeeds,
we neither need any information from $Cert$ nor do we need to propagate it to any other rule;
to keep this clear we use Prolog's anonymous variable.

{\bf Example of Implementation Elements} -

To concretize these technical elements,
we examine a few stages in the process of checking the trivial Trace proof given from Fig. \ref{fig:trace-ex}.

\begin{lstlisting}[language=Prolog]
check(certRight([4, 3, 1, 2],
  chains([chain(5,[3,1],x(1)),  chain(6,[4,2,5],false)])),
  store([],[]),
  unfk([or(and(x(1),x(2)),or(and(not(x(1)),x(2)),
      or(and(not(x(1)),not(x(2))),and(x(1),not(x(2))))))])).
\end{lstlisting}

As mentioned, the initial form of the query
uses the $certRight$ predicate, which contains $DexList$ variable with indices with which each clause should be stored,
as well as the derived clauses wrapped in the predicates of the form $\chain{Index}{DecideList}{Formula}$.
The list variables $SL$ and $NL$ within the $\sstore{SL}{NL}$ predicate are both intially empty,
and the original theorem is negated. Because the infix-to-prefix translation is done recursively, the
reader will note that the order of the conjunctions has changed in a predictable manner. The order of their indices in the
$DexList$ variable is also adjusted to ensure they are assigned to the correct clauses.

This query will match with the
unfocused ``or" rule, which is applied to leave us with the same query as above, except we now have an unfocused set of
formulas, instead of a single long one:
\begin{lstlisting}[language=Prolog]
[and(x(1), x(2)), or(and(not(x(1)), x(2)),
  or(and(not(x(1)), not(x(2))), and(x(1), not(x(2)))))].
\end{lstlisting}

Now the ``store" rule is applicable, and will store the conjunction $\aand{x(1)}{x(2)}$ with index 4 in the $SL$ variable.
Our query is now:
\begin{lstlisting}[language=Prolog]
check(certRight([3, 1, 2],
  chains([chain(5, [3, 1], x(1)), chain(6, [4, 2, 5], false)])),
  store([(4, and(x(1), x(2)))], []),
  unfk([or(and(not(x(1)), x(2)),
    or(and(not(x(1)), not(x(2))), and(x(1), not(x(2)))))])).
\end{lstlisting}

Following this pattern, the ``or" and ``store" rules are successively applied until the unfocused set is empty, and
each original clause and index tuplet has been added to the $SL$ variable.
At this point the only applicable rule is the ``cut" rule. This will select the first chain from the $ChainList$ and
begin a new branch of the proof. The chain's clause is negated, and
set added to the new unfocused set.
The $Cert$ variable now pertains only to this chain, and is wrapped in the $\certLeft{DecideList}{Flag}$ predicate to make the distinction clear,
resulting in the new query:
 \begin{lstlisting}[language=Prolog]
 check(certLeft([3, 1], 1),
 store([(2, and(x(1), not(x(2)))),  (1, and(not(x(1)), not(x(2)))),
    (3, and(not(x(1)), x(2))),  (4, and(x(1), x(2)))], []),
 unfk([x(1)])).
 \end{lstlisting}

From here, the branch is followed until $x(1)$ can be derived from the clauses 3 and 1 (as is indeed the case).
When the checker has managed to apply ``init" on all leaves of the proof, the left branch has succeeded and the subsequent
steps in the ``cut" rule allow the right branch to continue checking other derived clauses.

If, instead, the unfocused theorem cannot be derived from clauses indicated in the $DecideList$, this indicates either
an error in the SAT solver's proof, or that the SAT problem was not UNSAT in the first place (the first condition may
obviously occur
independently of the second, but not vice versa).

\section{Experiments}
\label{sec:exps}
As we have discussed in the previous sections, our checker for Trace format proofs follows the \PLKF system very closely,
and can therefore be highly trusted. However, it only works for relatively small Traces, which result from a relatively small
SAT problems. The checker would be of greater practical interest if it could handle large problems in a reasonable amount of time.
Therefore, in this section we explore why the checker currently is not efficient for large problems, and present several potential
improvements to the checker as well as to SAT solvers that output Trace. Our experiments suggest
that including the right information in a Trace could significantly improve our proof checker's ability to efficiently
check large proofs with a high degree of trust.

\subsection{Source of Non-Determinism}
In the focused sequent calculus, most rules are fairly deterministic. The focused  ``or" ($\vee$) rule is nondeterministic, but due to
our choice of polarization of the formulas, it will never be applicable and we have not implemented it.
The ``cut" rule would also be subject to nondeterminism,
but instead we supply the formula to cut on from the Trace, so there is no search necessary.
The main source of nondeterminism
is the ``decide" rule, where the checker may have arbitrarily many options to decide on, and only one or a few of them may
advance the proof. We postulate that this is where most of the non-determinism comes into play. If we don't know the
correct order for the $DecideList$, the problem becomes one of proof search.
In this case, we are using an automated theorem prover to check another theorem prover.  It is a
smaller proof search than we would have without incorporating the Trace proof evidence, but a proof
search nonetheless. This process can certainly increase the trust in the results of the first prover,
but is not efficient enough to easily check the results of large SAT problems.

The readme file for \texttt{Booleforce}'s Trace states that neither the global order of chains, nor the local order
of antecedents in a chain are guaranteed \footnote{http://fmv.jku.at/tracecheck/README.tracecheck},
rather any necessary ordering is
left for the checker to resolve by search -- therefore, the current, unmodified Trace format does not allow a
consistently linear search if its antecedent lists are not reordered.

The checker is therefore expected to perform search across clauses as well as across
antecedent chains. Although the latter requirement is satisfied by our checker, the first is not.
However, the global clause order output by \texttt{Booleforce} is already appropriate for \PLKF so
this was not a problem.

On the other hand,
resolution proofs do not use positive and negative phases while \PLKF does, the antedecent order in a Trace
proof will rarely
happen to also be the ideal order for a \PLKF proof.

Our first experiment is a brief example of this problem.
While we can verify a small Trace proof regardless of the local order of its chains' antecedents, in our
second experiment we demonstrate that the order of
antecedents makes a significant difference in the amount of proof search necessary to derive the proof.

To illustrate the issue of local antecedent order, consider part of a proof branch corresponding to the example Trace in
Sec. \ref{sec:prelim-trace}.
In Fig. \ref{fig:ProofExample}, we see the most direct path to eventual success (application of the ``init" rule) on all
branches. In the ``decide" marked by *, deciding on $3$ rather than $-1$ would result in failure and backtracking.
And in the very first ``decide," we decide on $1$ rather than $3$. If, at this juncture,
we instead decide on $3$, one possible result is Fig. \ref{fig:ProofExampleFail}. In that case, the search will fail
when no applicable rule is found for a positive focused literal, and need to backtrack to the ``decide."

Despite the impending failure, we see in Fig. \ref{fig:ProofExampleFail} that the right branch still succeeds as it does
in \ref{fig:ProofExample}. The checker may have to go through this successful branch before getting to the failure of our left branch.
In this trivial example, going through this small branch before reaching failure and backtracking is a benign exercise.
However, even small real SAT problems may require following much deeper trees before reaching failure.
Antecedents not taken in the right order, therefore, are potentially very costly to the checker's proof search.
This is why we focus on the subject of antecedent list order in our research.

\begin{figure}[tb]
    \scriptsize
  \begin{prooftree}

\AxiomC{}
\RightLabel{\texttt{init}}
\UnaryInfC{$\Sync{(-1)}{\Theta,(-1,A),(\neg{B})}{{B}}$}
\AxiomC{}
\RightLabel{\texttt{init}}
\UnaryInfC{$\Sync{}{\Theta,(\neg{B}),(\neg{A})}{{A}}$}
\RightLabel{\texttt{decide}}
\UnaryInfC{$\Async{(-1)}{\Theta,(-1,A),(\neg{B}),(\neg{A})}{}$}
\RightLabel{\texttt{store}}
\UnaryInfC{$\Async{(-1)}{\Theta,(-1,A),(\neg{B})}{\neg{A}}$}
\RightLabel{\texttt{release}}
\UnaryInfC{$\Sync{(-1)}{\Theta,(-1,A),(\neg{B})}{\neg{A}}$}
\RightLabel{\texttt{and}}
\BinaryInfC{$\Sync{(-1)}{\Theta,(-1,A),(\neg{B})}{\neg{A}\wedge {B}}$}
\RightLabel{\texttt{decide}}
\UnaryInfC{$\Async{(3,-1)}{\Theta,(-1,A),(\neg{B})}{}$}
\RightLabel{\texttt{store}}
\UnaryInfC{$\Async{(3,-1)}{\Theta,(-1,A)}{\neg{B}}$}
\RightLabel{\texttt{release}}
\UnaryInfC{$\Sync{(3,-1)}{\Theta,(-1,A)}{\neg{B}}$}
\AxiomC{}
\RightLabel{\texttt{init}}
\UnaryInfC{$\Sync{(3)}{\Theta,(-1,A),(\neg{A})}{A}$}
\RightLabel{\texttt{decide*}}
\UnaryInfC{$\Async{(3,-1)}{\Theta,(-1,A),(\neg{A})}{}$}
\RightLabel{\texttt{store}}
\UnaryInfC{$\Async{(3,-1)}{\Theta,(-1,A)}{\neg{A}}$}
\RightLabel{\texttt{release}}
\UnaryInfC{$\Sync{(3,-1)}{\Theta,(-1,A)}{\neg{A}}$ }
\RightLabel{\texttt{and}}
\BinaryInfC{$\Sync{(3,-1)}{\Theta,(-1,A)}{\neg{A}\wedge \neg{B}}$}
\RightLabel{\texttt{decide}}
\UnaryInfC{$\Async{(1,3,-1)}{\Theta,(-1,A)}{}$}
\RightLabel{\texttt{store}}
\UnaryInfC{$\Async{(1,3)}{\Theta}{A}$}
\end{prooftree}

\caption{A portion of the left branch of the proof search after the first application of ``cut." The $\Theta$
here always denotes the set of clauses $(1,\neg{A} \wedgen \neg{B}), (2,A \wedgen \neg{B}),
(3,\neg{A} \wedgen B), (4,A \wedgen B)$. Note that when we store $A$, it is stored as a tuple with an index.
When we store $\neg{A}$, it does not retain an index since it will not be decided on. This corresponds to the storing
of a negative atom in the $NL$ list in our Prolog implementation.
The context variable, refered to as $\Xi$ in Fig. \ref{fig:lkf} or as $Cert$
in the Prolog code, changes.  }
\label{fig:ProofExample}
\end{figure}

\begin{figure}[tb]
  \begin{prooftree}

\AxiomC{\texttt{fail}}
\UnaryInfC{$\Sync{(1,-1)}{\Theta,(-1,A)}{{B}}$}

\AxiomC{}
\RightLabel{\texttt{init}}
\UnaryInfC{$\Sync{(1)}{\Theta,(-1,A),(\neg{A})}{A}$}
\RightLabel{\texttt{decide}}
\UnaryInfC{$\Async{(1,-1)}{\Theta,(-1,A),(\neg{A})}{}$}
\RightLabel{\texttt{store}}
\UnaryInfC{$\Async{(1,-1)}{\Theta,(-1,A)}{\neg{A}}$}
\RightLabel{\texttt{release}}
\UnaryInfC{$\Sync{(1,-1)}{\Theta,(-1,A)}{\neg{A}}$ }

\RightLabel{\texttt{and}}
\BinaryInfC{$\Sync{(1,-1)}{\Theta,(-1,A)}{\neg{A}\wedge {B}}$}
\RightLabel{\texttt{decide}}
\UnaryInfC{$\Async{(1,3,-1)}{\Theta,(-1,A)}{}$}
\RightLabel{\texttt{store}}
\UnaryInfC{$\Async{(1,3)}{\Theta}{A}$}
\end{prooftree}

\caption{The same proof tree as in Fig. \ref{fig:ProofExample}, but at the first ``decide," $3$ is selected instead
    of $-1$.
}
\label{fig:ProofExampleFail}
\end{figure}

\subsection{Experiment 1: Global order of chains}
Although the Trace format assumes the checker will take charge of global clause order and local antecedent order through
proof search, our checker does not support the former. This is due to our use of the \PLKF cut rule, which will
only succeed when the antecedents of the newly introduced lema have already been derived, and stored. Our checker moves linearly
through the derived clauses, checking that for each cut formula, it is derivable from the context.
Therefore if a derived clause relies on another derived clause, yet comes before, the search will fail.
With our checker, we can trivially swap the order of any derived clauses, and it will fail.
We should keep in mind, therefore, that for our checker to work, the SAT solver must
maintain the correct global order of derived clauses. The SAT solver we used to obtain Traces with, \texttt{Booleforce},
consistently gives correct global order, but this may not be the case for other solvers.

\subsection{Experiment 2: Local order of antecedents in a chain}
In this experiment, our goal was to record the effect of antecedent order on the amount of proof search necessary to
derive a proof, in other words, the amount of non-determinism in the problem. We focus on relatively small problems where
there is only one or few chains that are longer than the rest. In these cases, we can reasonably assume that the longest of these
chains is responsible for a large portion of the problem's nondeterminism. Thus we can test the same problem but with different orders
of that one chain. We use runtime as a proxy for nondeterminism.
If our hypothesis is correct, and antecedent order is critical to reducing nondeterminism, we should see that
 the checker terminates much more rapidly with some orders of the longest chan than with other orders.
On such problems, we would expect that an increase complexity translates into only a small (hopefully linear) increase in runtime.

Our method is fairly simple: we implemented some additional methods in the Python program that translates the Trace format to
Prolog. One method finds the derived clause with the longest list of antecedents, and another allows us to replace the antecedent
list with another list. The test identifies the longest antecedent list, sorts the list in ascending order, and creates a
lexicographic permutation generator. For each permutation of the list, the translator replaces the original antecedent list with
the new permutation of itself (all other chains are left as they were),
recreates the Prolog file, and records the runtime of the file (as well as the output of Prolog, true or false).
From these results, we can select the best and worst performances.

These results are displayed in figures Fig. \ref{fig:longWorst} through \ref{fig:longBestLog} given in the appendix.

There are a few things to note about these results before analyzing them. First, in order to test a large number of permutations,
we set a timeout that each test cannot exceed. This affects our data in a few ways, but does not create an effect that doesn't exist --
if anything, it understates the effect. For the purposes of statistical analysis, when a test times out, we assign it the timeout
as its runtime. Therefore, the average runtime figures are understated, as we do not know how long it may have taken to terminate
had the program been allowed to continue indefinitely. The worst runtimes are similarly understated. Lastly, in cases where more
than half the tests did not terminate, this can even underestimate the median runtime. In Fig. \ref{fig:performanceTable}, the tests
that didn't terminate are  after the runtime.

Secondly, in problems where there are more than 1000 permutations of the longest chain, we don't test all permutations but
rather test a random sample of all permutations (generated by a random shuffle). Again, if anything this understates the effect of
antecedent order, as the best and worst orders may not have been tested. Even in these cases, there is still a significant effect
 in only the permutations tested.

These data show not only an effect of antecedent order on runtime (and thereby the amount of proof search necessary),
but they also suggest that the Trace checking could be much more scalable to large problems if the largest chains come with a correct
or even nearly-correct order. This effect would be magnified, we propose, if all chains could have the right order (this is
explored in the next experiment).
Basic statistical analysis, which of course cannot be heavily relied upon in a case with
a small sample size, suggests that the model that best explains variation (with an R squared of 70\%) in complexity vs. worst
runtime is exponential. On the other hand, the model that best explains variation (however, this
time with an R squared of only 25\%) in complexity vs. best runtime is
a power model. If all chains were in the correct order, we would expect to see a very linear relationship between complexity
and best runtime. As it is in this experiment, the order of the longest chain clearly reduces the runtime significantly,
but other unordered
chains still require some level of proof search.

The low explanatory power of complexity in both these cases is certainly partly due to
the fact that the longest chain length is not a perfect proxy for complexity. We have tried to select problems for
this portion of the experiment where the resulting Trace had only one or few long chains, but different problems have varying numbers of
medium or longer chains in addition to the longest chain, which impacts their actual complexity but not the proxy. For this
reason we have also included average and median chain lengths for consideration. As an example, in Fig. \ref{fig:performanceTable}
we might consider the problem ``rksat13" somewhat of an outlier because, while the longest chain has 8 antecedents,
there are enough chains with between 6 and
8 antecedents to raise the median chain length to 6. We can reasonably assume that this is one reason why ``rksat13"
has slower runtimes than other problems with comparable longest chain lengths.
Therefore, we also include Fig. \ref{fig:medBestWorst} which uses median chain length as the indicator for complexity. We see
that this, too, has some correlation.

Finally, it is worthwhile to note that we have also preserved data from this experiment detailing which antecedent orders correspond
to which runtimes. This might be interesting for further analysis to establish a pattern or heuristic for ordering antecedent lists.

\subsection{Experiment 3: Checker that succeeds only when no proof search is necessary}
In this experiment, we adjusted the Prolog checker code so that it could not conduct any proof search along
an antecedent list. That is to say,
if in any chain the checker decides on an antecedent index that leads the proof to a failed branch, the checker cannot simply backtrack
and try deciding on a different index. In this way, if the proof does succeed, it means that there was no proof search necessary
across local antedecent lists,
and that the antecedents of all chains were in the correct order.
It should be noted that at this point, we do not know what the correct order for all
the clauses may be in any given example; we only know that in most cases, a resolution proof is unlikely to provide the ideal order
for a nearly deterministic \PLKF search, because resolution proofs do not have negative and positive phases.
Therefore the correct way to reorder clauses before passing them to the checker is another question of interest.

To do eliminate proof search in antedecent lists, we just adjusted the decide expert so that only the head of the $DecideList$ can be selected, as follows:

\begin{lstlisting}[language=Prolog]
%decide expert
decidee(certLeft(1,DL), certLeft(0,DL), -1).
decidee(certLeft(Flag,[I|Rest]), certLeft(Flag,Rest), I).
\end{lstlisting}

Testing this code requires an exhaustive enumeration of all permutations for each chain, as well as all combinations of each of these
permutations. The number of all such combinations for a Trace with $n$ chains, each with variable length $c_{i}$ is:

$c_{1}! \cdot c_{2}! \cdot c_{3}! \cdots c_{n}!$.

As a result, we were only able to test our smaller samples, as we were working with a Cloud9\footnote{\url{https://c9.io/}} instance with limited resources.

The problems we successfully used for this experiment were: madeUp, madeup2, simpleTrace, readme, and randKsat2;
their details can be found in \ref{fig:performanceTable}.
However, in each example we can observe that all combinations of permutations fail rapidly except for a few (between
one and six), which succeed rapidly. Although
we lack data on larger problems, from the sample we do have, it appears that the runtime has a linear relationship with problem
complexity, given that the Trace
has the optimal order
for every chains' antecedent lists. Therefore we propose that the correct order of the antecedents list
can lead to nearly full determinism in our proof certification and therefore scale to the checking of very large proofs.

Assuming every derived clause's antecedent list is ideally ordered for \PLKF proof search, is there perfect determinism?
Not quite, for at some point, a positive literal stored with $-1$ will need to be selected before the checker can apply the ``init"
rule. Even perfect antecedent order does not advise as to which literal should be chosen, if there are more than one marked by $-1$.
Furthermore,
in some cases it is
 necessary to decide on $-1$ before the DecideList is empty. However, testing the $-1$ option is very cheap
because the selection of a stored positive literal is only correct if it leads directly to application of ``init."
If this occurs,
the branch succeeds, otherwise it fails and backtracks quickly.
So given perfect antecedent list order,
the remaining nondeterminism is very shallow and is a function of the number of antecedents in each clause and the
number of literals stored with $-1$ at each step in the proof.

Again, as in the last experiment, we do retain the information for which combination of permutations succeeded, so this
would also be an interesting subject for further analysis.

\section{Conclusion and Future Work}
\label{sec:conc}
In this paper, we considered the problem of certifying very large unsatisfiability proofs with a high degree of trust.
Our approach was to write a certifier based on the sequent calculus, a low level-proof system, in the programming
language Prolog, which supplies us with unification and backtracking needed for conducting proof search.
The use of the sequent calculus allows any programmer with basic understanding in logic to implement our kernel.
The use of a logic programming language allows us to separate the implementation of the above algorithms from the amount
of trust we can place in the result. The reason for that is the existence of many implementations of Prolog,
such as SWI-Prolog \footnote{\url{http://www.swi-prolog.org/}} and GNU Prolog \footnote{\url{http://www.gprolog.org/}} which
we can use in order to execute the kernel program and, therefore, decrease the chance of a bug in those algorithms.

Our implementation clearly shows that our approach does not scale to big proofs when using the Trace format as proof evidence.
The experiments conducted imply that the reason for that is the lack of order between antecedents and chains in the proof evidence.
It first shows that our approach does not work at all given incorrect chain orders and we require the input chains to be correctly
sorted. Sorted chains is the behavior exhibited by the prover - \texttt{Booleforce} - we used in order to generate our examples.
We therefore require, for our approach to work, that this order be restricted in the evidence.

A more interesting source of complexity comes from the arbitrary orders of antecedents within the chains. We first demonstrate
the significant gap in performance as exhibited by the best and worst orders. We follow this experiment by
showing the possible scaling of our approach to more complex problems, given a correct order of antecedents.

One way of obtaining traces of correct order is to use a theorem prover to post-process the traces obtained from TraceCheck.
Alternatively, one can require SAT solvers to produce  ``ordered" traces. While the latter method produces more adequate proofs, it
is clearly irrelevant for determining the level of trust we can place in the results of SAT solvers.

Since our experiments required the generation of many permutations, we could not extend them to problems of real complexity. We hope that
our prototype prover shows the possibility of using the ProofCert ideas in order to certify SAT refutations.
 At the same time, we would like to improve our implementation to deal with problems of arbitrary
complexity (given a correct order). To do so, we intend on changing the main data structures of our program to use Prolog's internal ones.
More specifically, we will on replace data structures based on lists and their associated relations for testing for membership with
language level predicates and meta relations for testing the existence of such a predicate. Luckily, this type of support exists
in some Prolog distributions under meta relations such as \texttt{assert} and \texttt{retract}.

{\bf Acknowledgments.} We would like to thank the anonymous reviewers for the many constructive remarks.

\nocite{*}
\bibliographystyle{eptcs}
\bibliography{generic}

\appendix
\subsection{Graphical Results of the Experiments}
\label{sec:apx}
\begin{figure}[H]
  \center
  \includegraphics[width=0.8\linewidth]{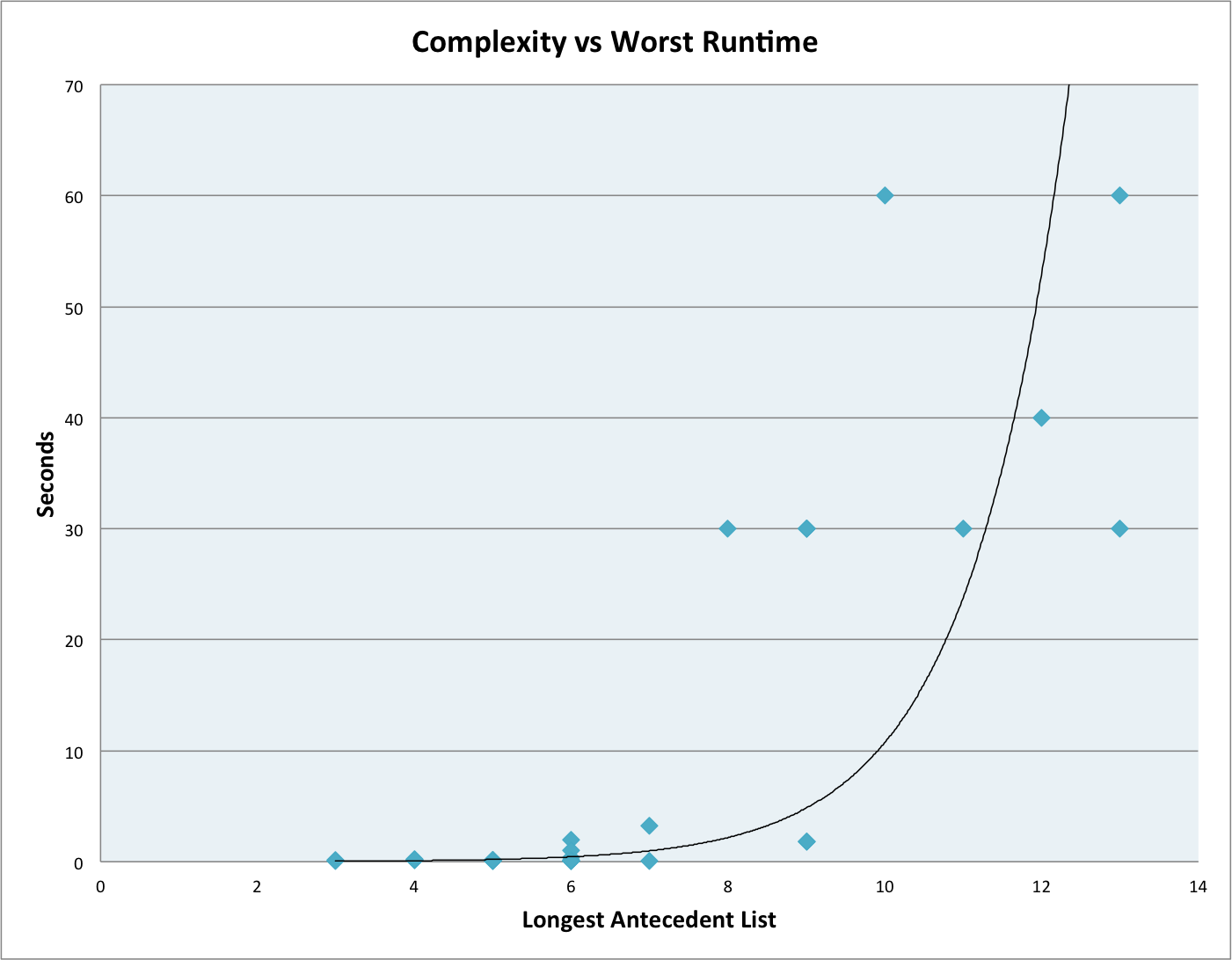}
  \caption{ Longest chain length vs. worst runtime in seconds (due to incorrect order of longest chain).}
  \label{fig:longWorst}
\end{figure}

\begin{figure}[H]
  \center
  \includegraphics[width=0.8\linewidth]{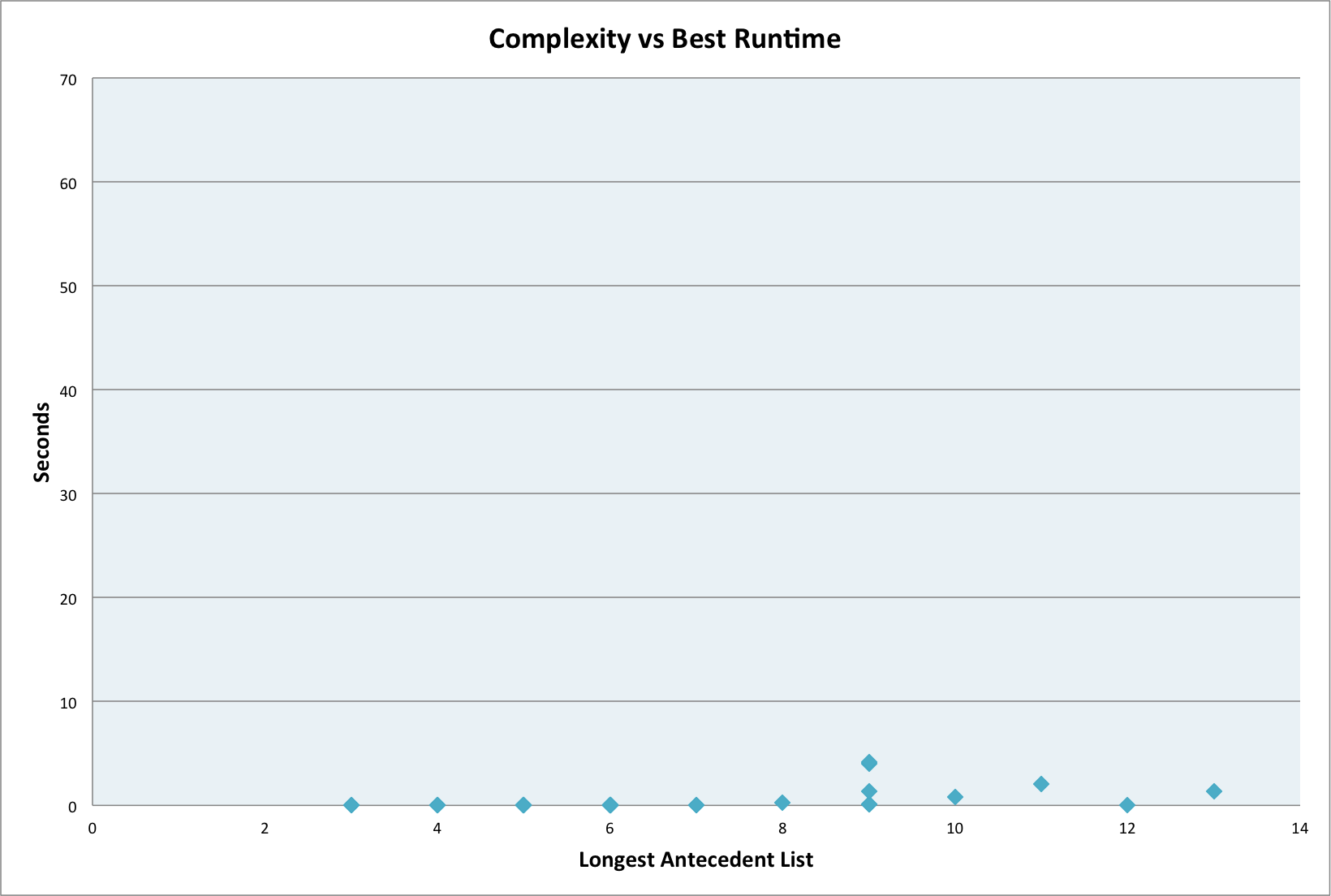}
  \caption{ Longest chain length vs. best runtime in seconds (due to correct order of longest chain); scale left the same as the previous graph for comparison.}
  \label{fig:longBest}
\end{figure}

\begin{figure}[H]
  \center
  \includegraphics[width=0.8\linewidth]{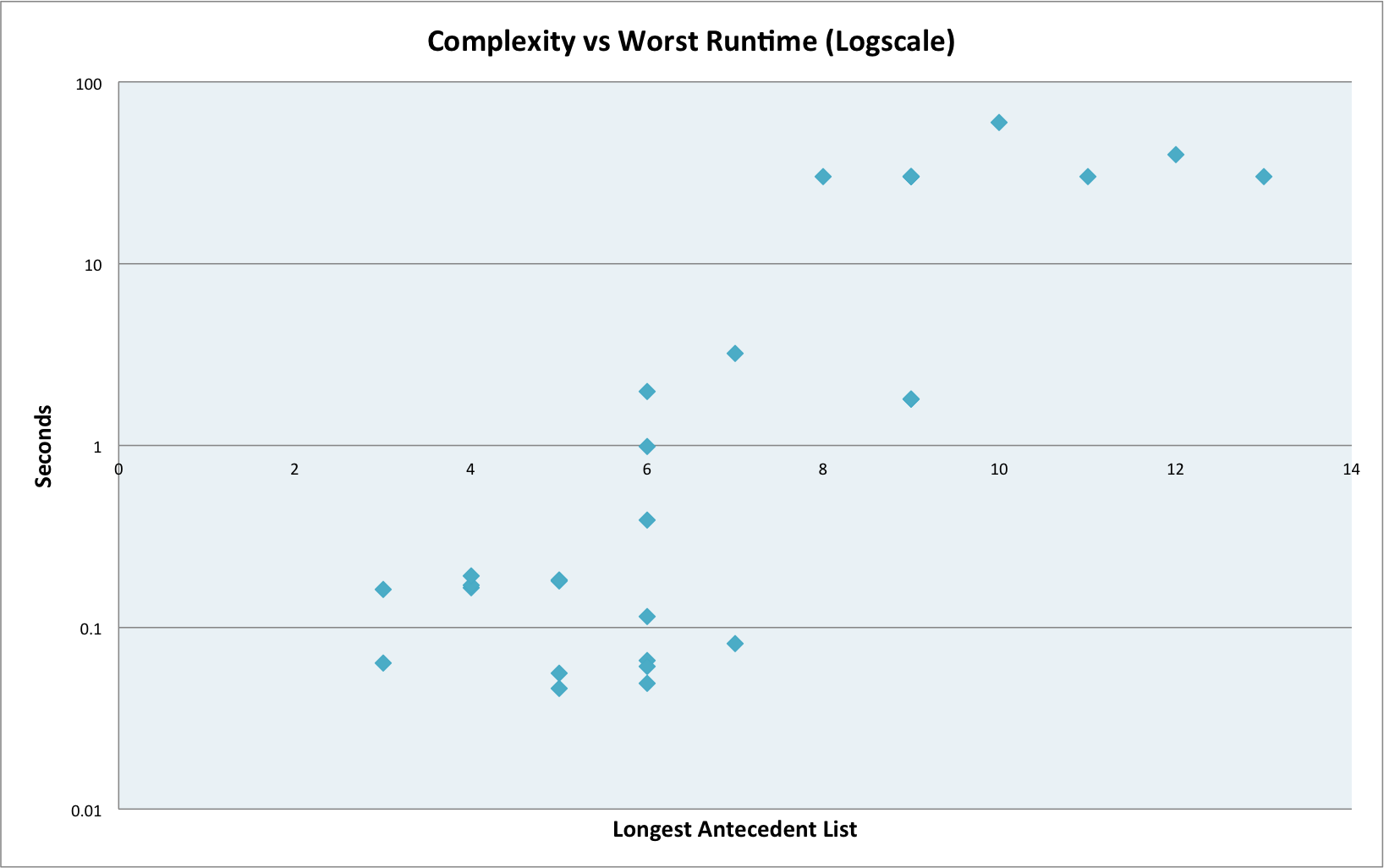}
  \caption{Fig. \ref{fig:longWorst} where the y-axis is logscaled to highlight pattern in very small changes.}
  \label{fig:longWorstLog}
\end{figure}

\begin{figure}[H]
  \center
  \includegraphics[width=0.8\linewidth]{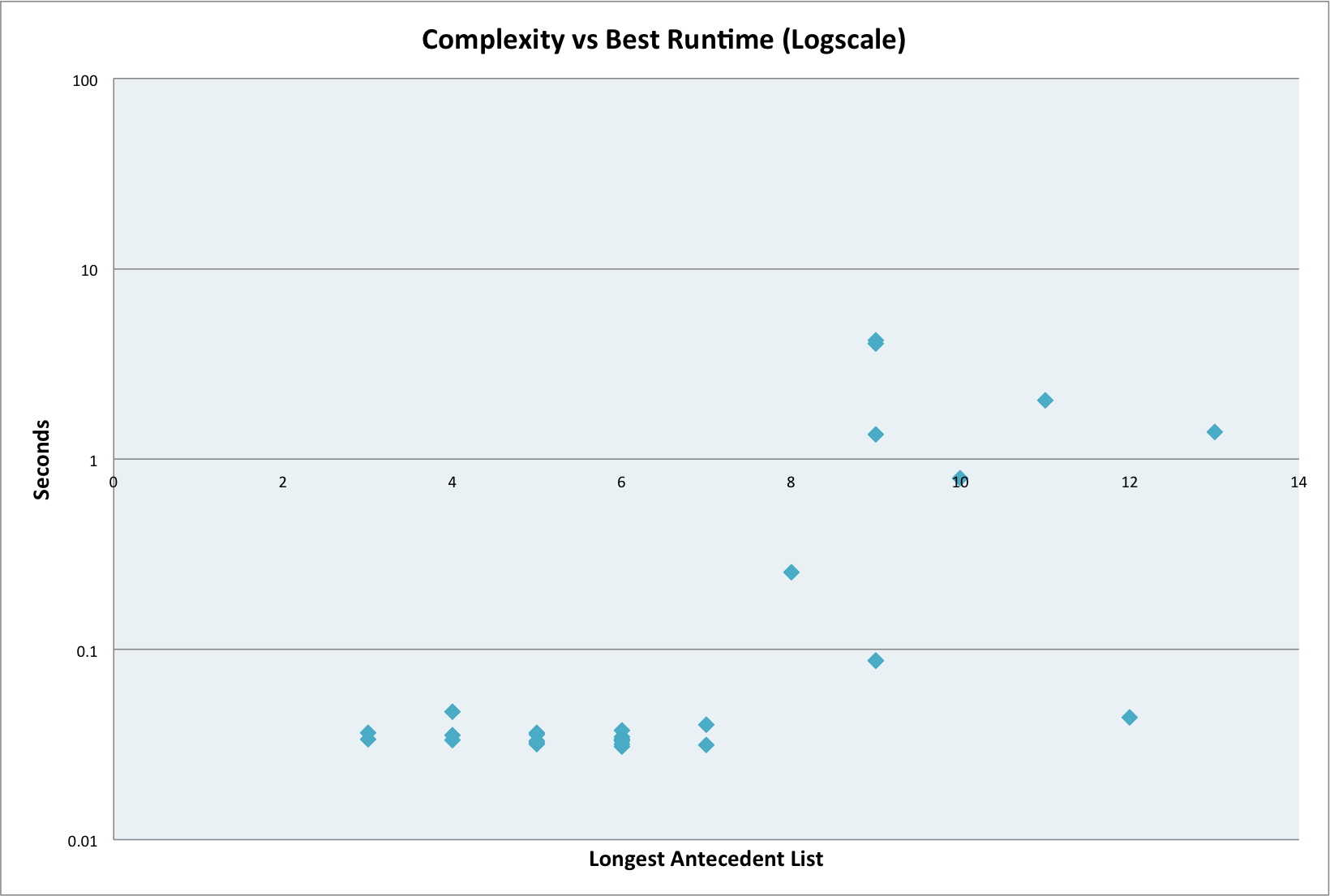}
  \caption{Fig. \ref{fig:longBest} where the y-axis is logscaled to highlight pattern in very small changes.}
  \label{fig:longBestLog}
\end{figure}

\begin{figure}[H]
  \center
  \includegraphics[width=0.8\linewidth]{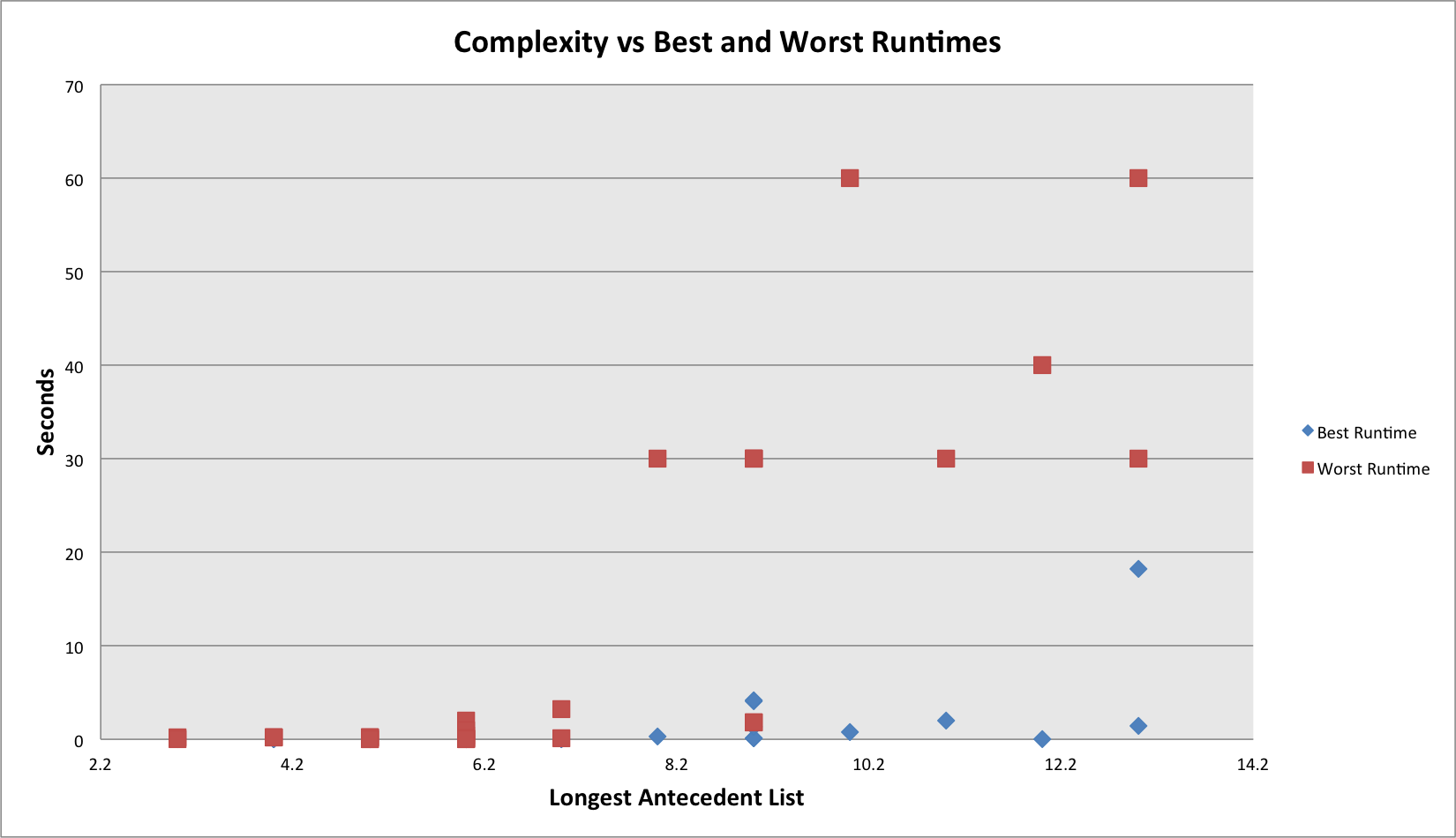}
  \caption{ Longest chain length vs. best and worst runtimes (due to reordering the longest chain).}
  \label{fig:longBestWorst}
\end{figure}

\begin{figure}[H]
  \center
  \includegraphics[width=0.8\linewidth]{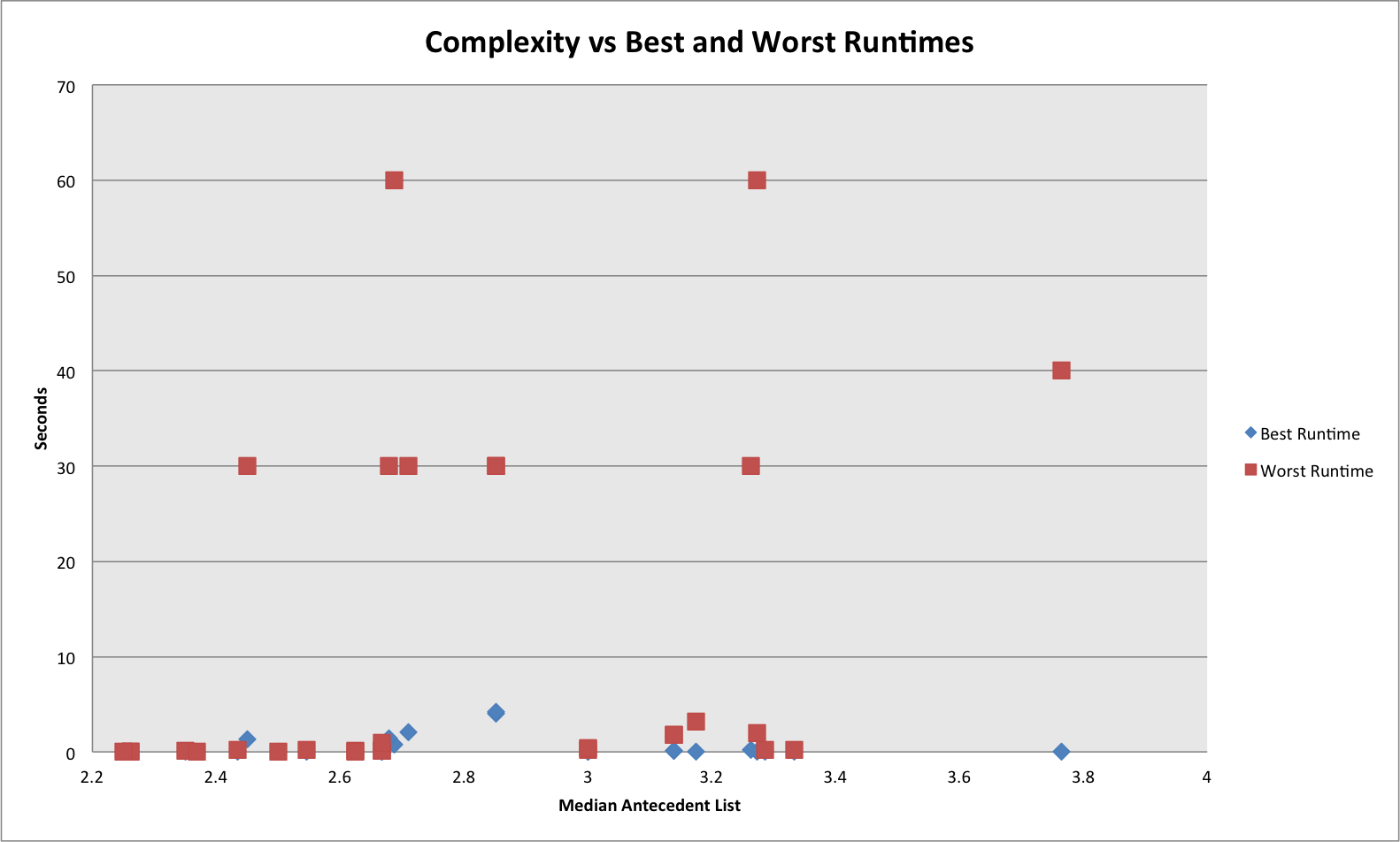}
  \caption{ Median chain length vs. best and worst runtimes (due to reordering the longest chain).}
  \label{fig:medBestWorst}
\end{figure}

\begin{figure}[H]
  \center
  \includegraphics[width=0.9\linewidth]{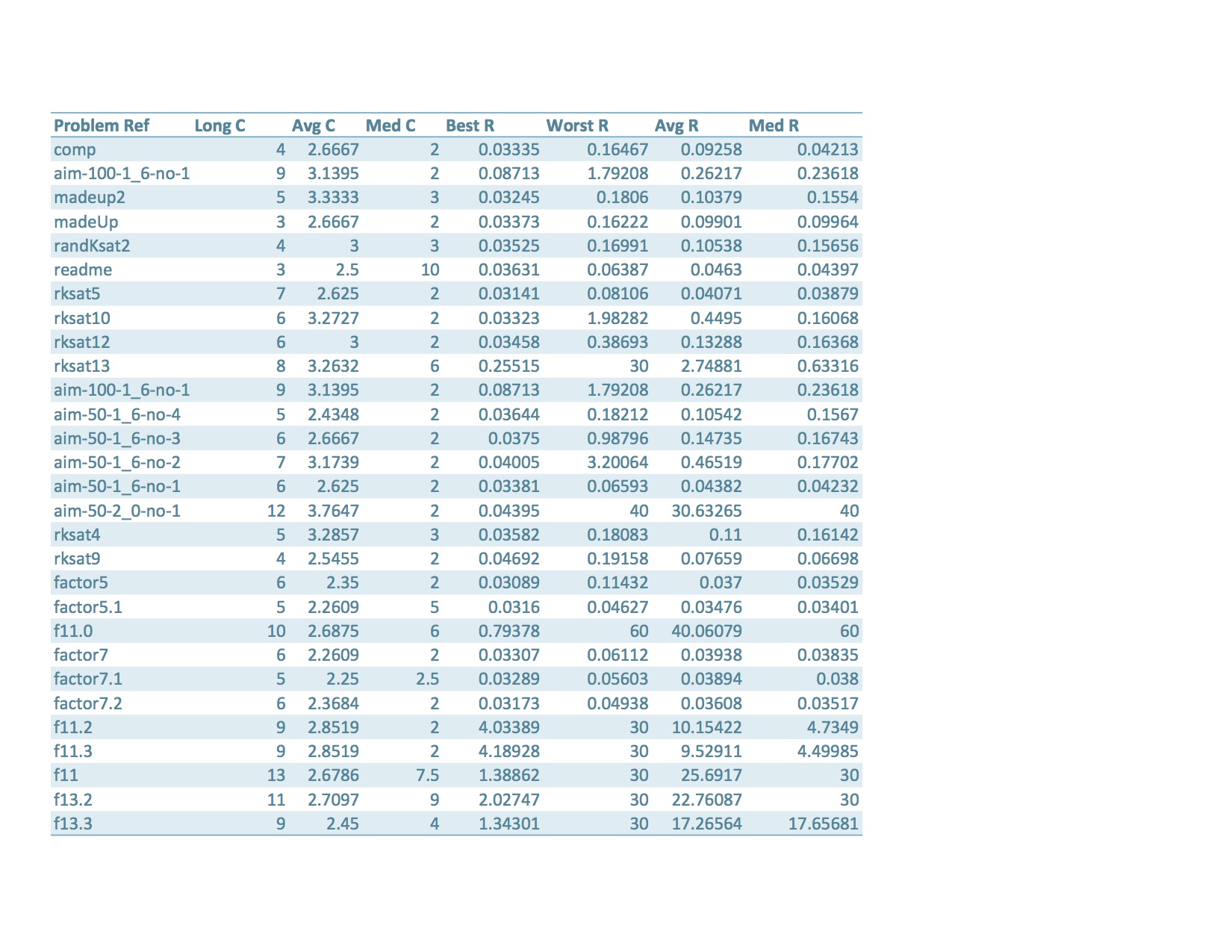}
  \caption{Data summary. 'C' stands for chain, 'R' stands for runtime, 'avg' and 'med' stand for average and median.}
  \label{fig:performanceTable}
\end{figure}

\end{document}